\documentclass [twoside]{article}
\usepackage{epsfig}
\author{ P. Kulinich \\
\small MIT, Cambridge, MA, USA;  \ \ kulinich@mit.edu
}

\title{
Novel String Banana Template Method of Track Reconstruction for high 
 Multiplicity Events with Significant Multiple Scattering
}

\begin{document}
\markboth{P. Kulinich}{Tracking with significant MS}

\maketitle

\begin{abstract}
Novel String Banana Template Method (SBTM) for track reconstruction in 
high multiplicity events in non-uniform magnetic field spectrometer with 
emphasis on the lowest momenta tracks with significant Multiple Scattering
(MS)  is described.
Two steps model of track with additional parameter/s which takes into 
account MS for this particular track is introduced.
SBTM is time efficient and demonstrates better resolutions
than another  method equivalent to the Least Squares method (LSM).  
%\keywords{Track; Reconstruction; Multiple scattering.}
\end{abstract}

\section{SBTM description}

 Detailed Monte Carlo (MC) simulations 
(using GEANT \cite{GEANT_}, for example) could provide database 
of track's characteristics  (templates) for later fast use.

\begin{figure}  [h]
\begin{center}
\mbox{\epsfig{bbllx=0mm,bblly=1mm,bburx=150mm,bbury=120mm,%
%height=5cm,width=5cm,file=SBTM_fig1_1.eps,clip=}%
height=6cm,width=6cm,file=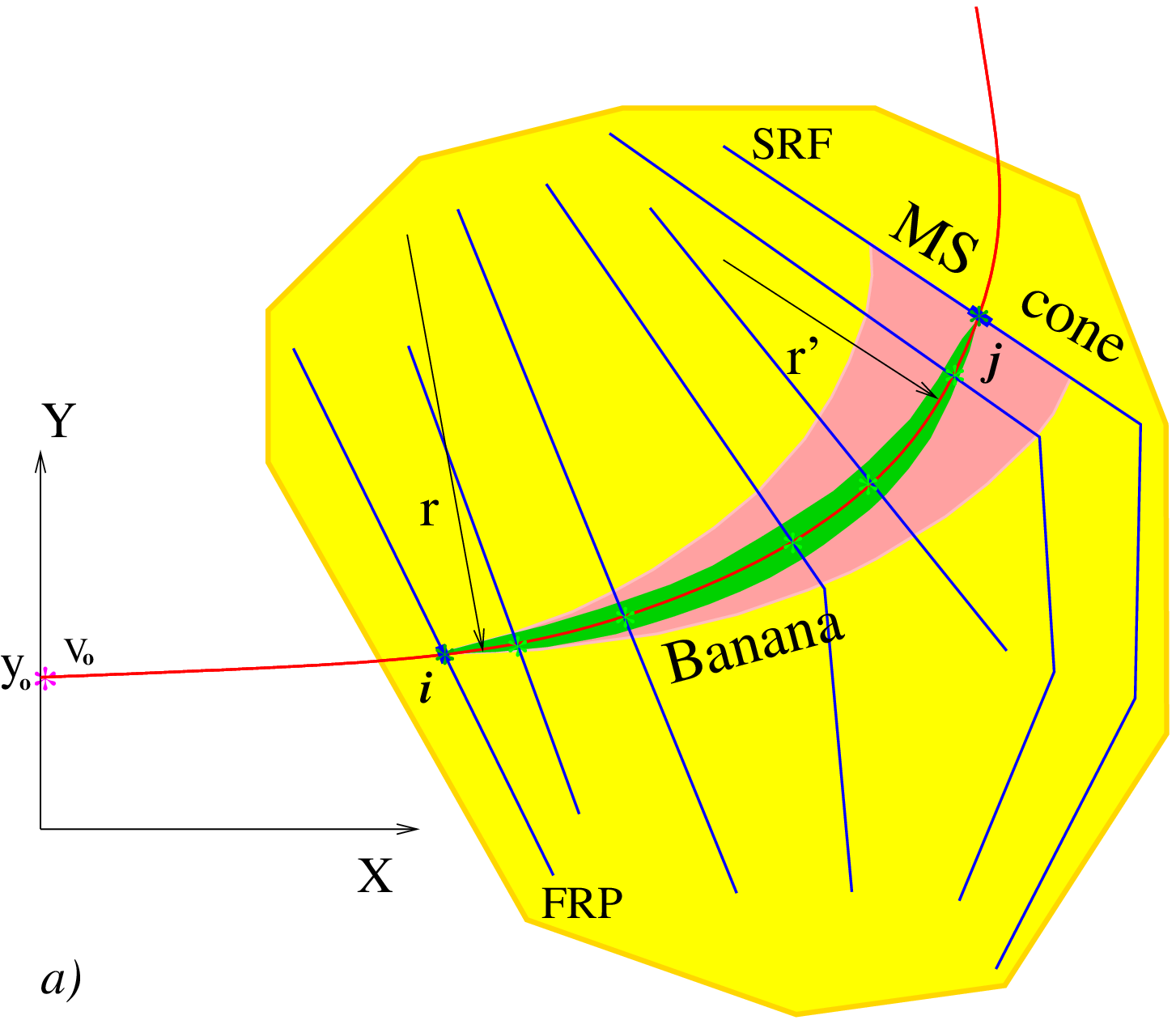,clip=}%
}
\mbox{\epsfig{bbllx=5mm,bblly=0mm,bburx=194mm,bbury=270mm,%
%height=9.5cm,width=7.5cm,file=SBTM_fig1_2.eps,clip=}%
height=11.5cm,width=8.7cm,file=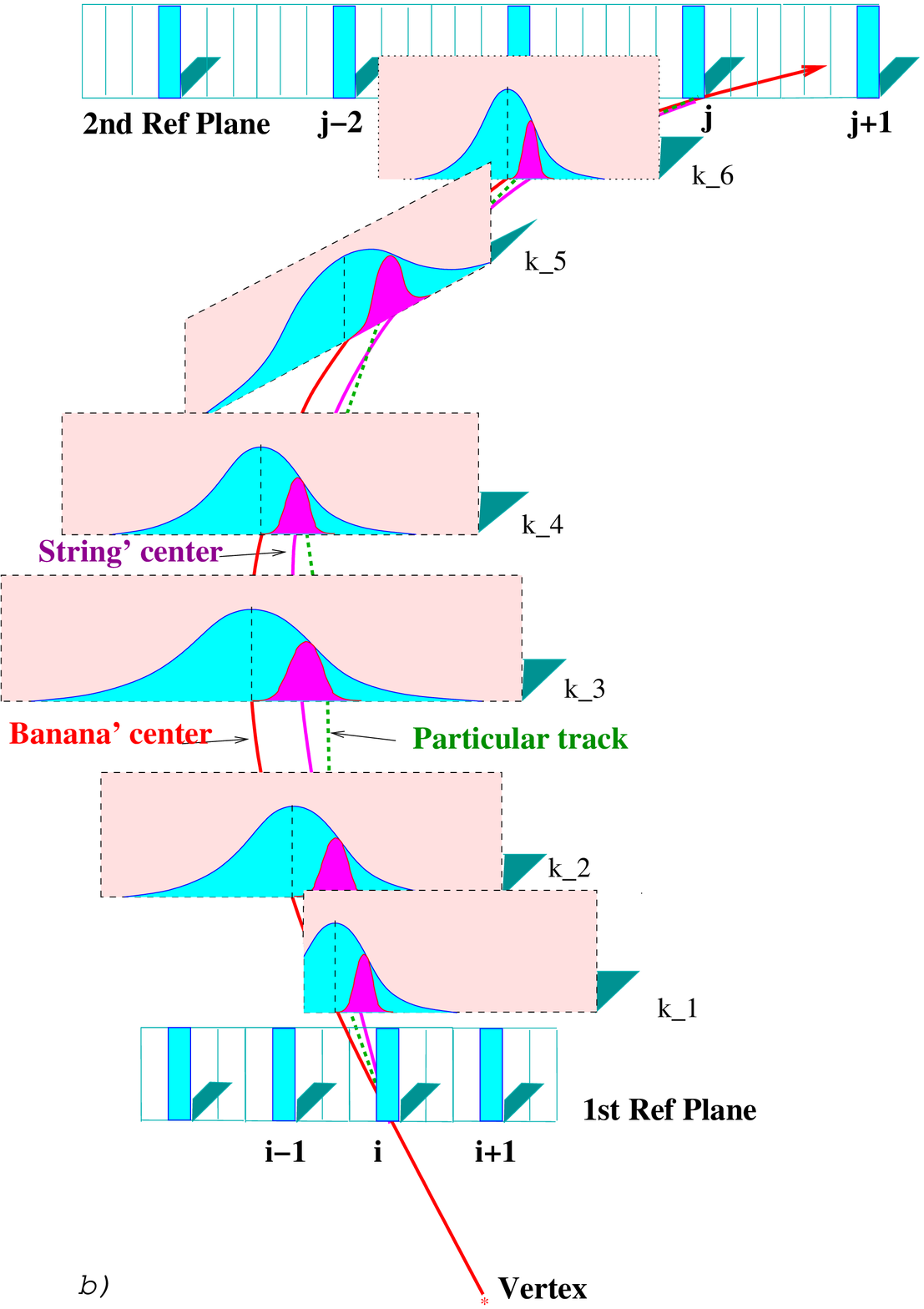,clip=}%
}
\end{center}
\vspace*{8pt}
\caption{$a)$ MS cone for ensemble of particles with 
monochrome  $\vec P$ and Banana 
for ensemble with the same average but narrow momentum 
distribution, which  originated from $V_{0}$ and
pass  through the same pixel $i$ in FRP. 
$b)$ Banana $(i,j)$ ``bell road'' and a particular String 
(more narrow ``bells'') inside.}
\label{SBTM_log}
\end{figure}

The MS cone (as on Fig.~\ref{SBTM_log}$a)$) for an ensemble of particles 
with the same vector momentum $\vec{P}$ has a significant width (volume) 
at low momenta. At fitting stage in high multiplicity event it results
 in heavy computation with  covariance matrices \cite{Grote_}.
For pattern recognition in such difficult cases it's crucial to have 
narrow Search Windows (SW)  what requires an  a priori
knowledge of momentum.

%\end{document}

It was  found  \cite{SBTM_1conf} that for another ensembles SWs are more narrow.
The main idea of SBTM is in use of ensembles 
with 3 fixed points. Where the 
first  point is  initial vertex $V_{0}$. As two other points 
 two strips (pixels) $i,j$ in two reference planes (RP) close to the middle
and end of the track are used. It results in 2-dimensional 
 $\{i,j\}$ templates (see  Fig.~\ref{SBTM_log}$b)$). 
For such ensemble which geometrical image in magnetic field has 
a $Banana$-like shape  all necessary characteristics 
(centre of Bananas, their widths, angles, lengths ... ) 
could be saved a priori to track reconstruction.  

 Two steps model of track is used: the first  one -- averaged over ensemble, 
 gives rough estimations for SW ($Banana$ center and width) 
and momentum of candidate.  The second -- per event dependent 
(takes into account MS for this particular track), permits to localize 
track in more narrow $String$ SW and provides 
track parameters corrections depending on relative deflection of a $String$ 
inside a $Banana$ window. 
So SBTM uses more parameters in its model space than usually and it
 provides additional corrections to track' parameters.

 At first track recognition stage one check
 different combinations of $(i,j)$
 signals in RPs and select such which has proper number of signals 
 inside of $Banana$ road. At the second stage 
relative positions of signals inside $Banana$ are checked and if they are 
inside of a more narrow window $String$ -- track candidate is recognized.

\section{Comparison to another method and Conclusions}

For demonstration and comparison the SBTM method is applied for the toy model
 spectrometer as in \cite{Lutz_}. In this article  optimal track fitting (OTF)
 which reproduces the results of the global fit \cite {Wolin_} is described. 
Toy model spectrometer consists of four high resolution (5 $\mu$m)
silicon detectors  followed by thirteen gas detectors 
(200 $\mu$m resolution) in 1T magnetic field.

\begin{figure} [ht]
\begin{center}
%\mbox{\epsfig{file=SBTM_fig2.eps,height=8.8cm,width=12cm,% 
\mbox{\epsfig{file=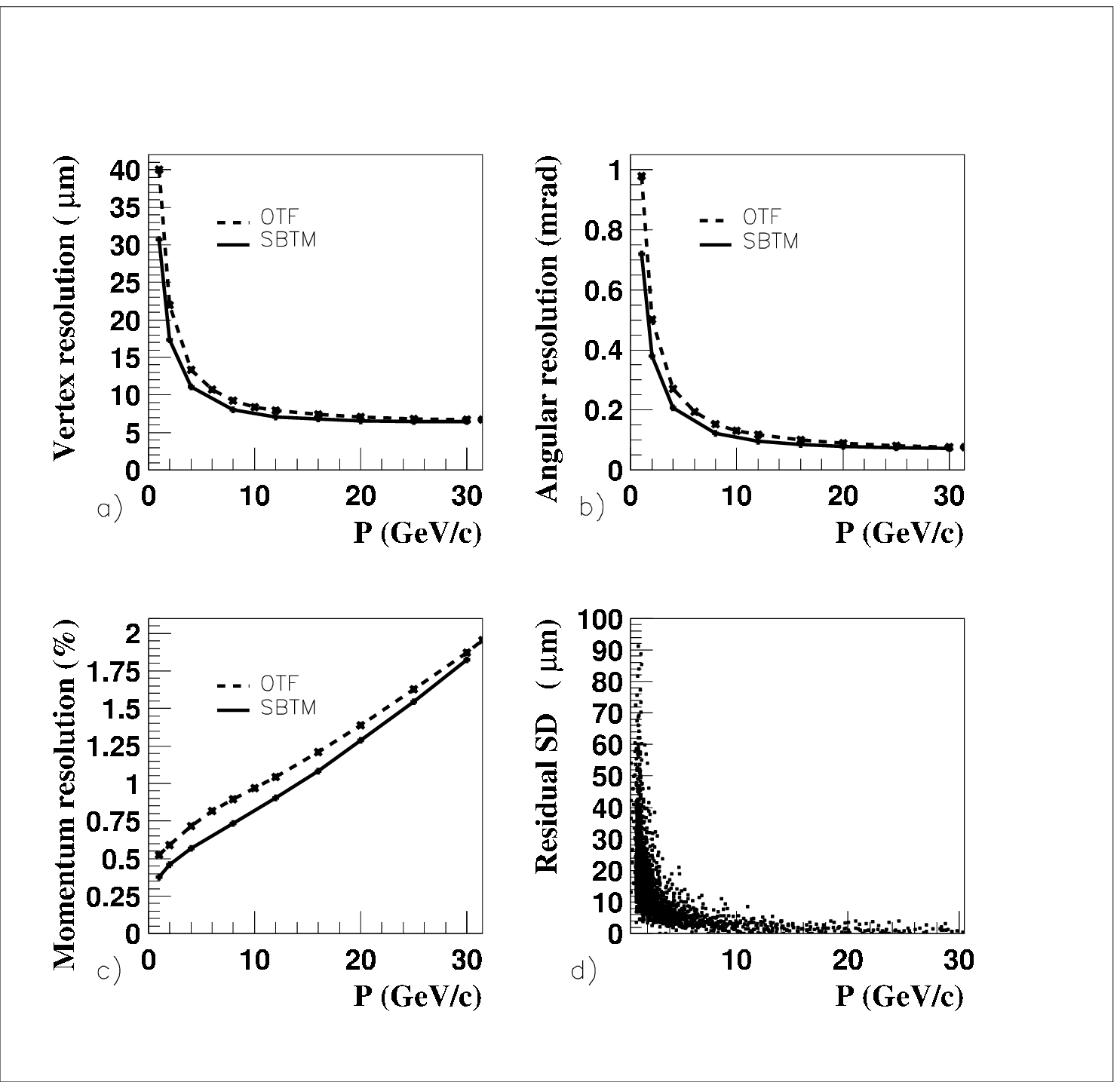,height=11cm,width=14.5cm,% 
bbllx=5mm,bblly=5mm,bburx=140mm,bbury=130mm,clip=}%
}   
\end{center}
\vspace*{8pt}
\caption{{\small Vertex $a)$, angular $b)$  and momentum $c)$ resolutions  as a 
function of particle  momentum. Space precision of the SBTM model --
 residual standard deviation (for ``ideal'' space resolution case).}}
\label{Imp_par}
\end{figure}

 Main track reconstruction characteristics for this method and for
 OTF are shown on Fig.~ \ref{Imp_par}.
 Points for OTF were taken from Fig.~10,11,12 
 in \cite{Lutz_}. 
 Fig.~\ref{Imp_par}d) shows how close is track model to the actual hits.

This  global method  has internal robustness and can easily work with
ambiguous measurements of different detectors.
It exploits simple and fast access model of track and is time
 efficient. Its template based nature and close approach to the actual hits 
make it attractive for implementation in firmware.

\section*{Acknowledgments}
 The author is grateful to W. Busza for support.
This work was partially supported by U.S. DoE grant
DE-FC02-94ER40818.

\end{document}